\documentclass[journal]{IEEEtran}

\usepackage{xcolor,soul,framed} %,caption
\usepackage{mathrsfs}
\usepackage{dsfont}
\usepackage{amsfonts}
\usepackage{changepage}
\usepackage{bm}
\usepackage{booktabs}
\usepackage{multirow}
\usepackage{makecell}
\usepackage[ruled,linesnumbered]{algorithm2e}
\usepackage{amssymb}
\usepackage[hyphens]{url}
\usepackage[breaklinks, colorlinks, linkcolor=purple, anchorcolor=purple, citecolor=purple, urlcolor=purple]{hyperref}

\colorlet{shadecolor}{yellow}
\usepackage[pdftex]{graphicx}
\graphicspath{{../pdf/}{../jpeg/}}
\DeclareGraphicsExtensions{.pdf,.jpeg,.png}

\usepackage[cmex10]{amsmath}
\usepackage{array}
\usepackage{mdwmath}
\usepackage{mdwtab}
\usepackage{eqparbox}
\usepackage{url}

\begin{document}
\bstctlcite{IEEEexample:BSTcontrol}
    \title{An Adaptive Approach for Probabilistic Wind Power Forecasting Based on Meta-Learning}
  \author{Zichao Meng,~\IEEEmembership{Student Member,~IEEE,}
      Ye Guo,~\IEEEmembership{Senior Member,~IEEE,}
      and~Hongbin Sun,~\IEEEmembership{Fellow,~IEEE}% <-this % stops a space% <-this % stops a space
      
  \vspace{-0.5cm}
  \thanks{ This work was supported in part by the National Key R\&D Program of China under Grant 2020YFB0906000 and 2020YFB0906005. Corresponding author: Ye Guo, e-mail: guo-ye@sz.tsinghua.edu.cn}
  \thanks{Z. Meng and Y. Guo are with the Smart Grid and Renewable Energy Lab, Tsinghua-Berkeley Shenzhen Institute (TBSI), Tsinghua University, Shenzhen 518055, China.
  }% <-this % stops a space
  \thanks{H. Sun is with the Department of Electrical Engineering, State Key Laboratory of Power Systems, Tsinghua University, Beijing 100084, China, and also with the Smart Grid and Renewable Energy Lab, Tsinghua-Berkeley Shenzhen Institute (TBSI), Tsinghua University, Shenzhen 518055, China.}%
  }

% \markboth{IEEE TRANSACTIONS ON SUSTAINABLE ENERGY}{Roberg \MakeLowercase{\textit{et al.}}: Meta-learning Learning}

% ====================================================================
\maketitle

% === ABSTRACT ====================================================================
% =================================================================================
\begin{abstract}
%\boldmath
This paper studies an adaptive approach for probabilistic wind power forecasting (WPF) including offline and online learning procedures. In the offline learning stage, a base forecast model is trained via inner and outer loop updates of meta-learning, which endows the base forecast model with excellent adaptability to different forecast tasks, i.e., probabilistic WPF with different lead times or locations. In the online learning stage, the base forecast model is applied to online forecasting combined with incremental learning techniques. On this basis, the online forecast takes full advantage of recent information and the adaptability of the base forecast model. Two applications are developed based on our proposed approach concerning forecasting with different lead times (temporal adaptation) and forecasting for newly established wind farms (spatial adaptation), respectively. Numerical tests were conducted on real-world wind power data sets. Simulation results validate the advantages in adaptivity of the proposed methods compared with existing alternatives.
\end{abstract}

% === KEYWORDS ====================================================================
% =================================================================================
\vspace{-0.2cm}
\begin{IEEEkeywords}
Wind power forecasting (WPF), probabilistic forecasting, temporal and spatial adaptabilities, meta-learning, online learning, quantile regression 
\end{IEEEkeywords}

% \fi
\IEEEpeerreviewmaketitle

% ====================================================================
% ====================================================================
% ====================================================================

% === I. INTRODUCTION =============================================================
% =================================================================================
\vspace{-0.3cm}
\section{Introduction}

The development of wind energies is of crucial importance for carbon neutrality. Until 2021, the total worldwide installed capacity of wind energies has reached over 823 GW \cite{IRENA-2022}, and higher penetration levels are expected. With the rapid development of wind power generation resources, power systems are facing increasing uncertainties on the generation side. In practice, dispatch centers recursively schedule the look-ahead dispatch based on the forecast of wind generation outputs. Prominent wind power forecasting (WPF) techniques are thus imperative, which help to optimize decision-making with different lead times. 

Owing to the more flexible implementations and excellent nonlinear approximation ability compared with physical model-based methods \cite{Landberg-WE-2025} and conventional statistical methods \cite{Zhao-TPS-2018}, machine learning methods \cite{Buhan-TII-2015}-\cite{Wang-TSE-2021} are widely used in WPF. Existing machine-learning-based WPF methods can be categorized into two classes: point forecast and probabilistic forecast \cite{AEE-2021}. Examples of point forecast models include support vector machines \cite{Buhan-TII-2015}, extended polynomial networks \cite{Zjavka-TSE-2018}, and Markov chains \cite{Sanjari-TSE-2020}. Probabilistic WPF methods offer more information (quantiles, intervals, or densities) \cite{Wen-TSE-2022}-\cite{Wang-TSE-2021} compared with the point ones (only expectations) \cite{Buhan-TII-2015}-\cite{Sanjari-TSE-2020}, which is of crucial importance for the decision-making of power system operators with extensive penetration of renewable energies. For probabilistic forecast models, representative methods include Gaussian process \cite{Wen-TSE-2022}, nonparametric Bayesian method \cite{Xie-TPS-2019}, and kernel density estimation \cite{Wang-TSE-2021}. Recently, learning algorithms with deep architectures have also been adopted to further improve the forecasting performance, and recurrent neural network (RNN)-based models have been adopted in these researches \cite{Shi-TSE-2018}. In \cite{Ko-TSE-2021}, the bidirectional long short-term memory (LSTM) network with deep concatenated residual structure was used for WPF. In \cite{Wang-TSE-2022} and \cite{Arora-TII-early}, the Bayesian LSTM and the autoregressive RNN were proposed for the probabilistic WPF. Besides, deep mixture density network \cite{Zhang-TPS-2020}, stacked autoencoder \cite{Chen-TSE-2020}, ensemble convolutional neural network (CNN) \cite{Wang-AE-2017}, and temporal attention network \cite{Zhang-TSE-2021} were applied to WPF.

%Although aforementioned approaches show prominent performance in both the point and probabilistic WPF, they assume there is a large amount of data to train the forecast model in an offline learning manner. In other words, these approaches do not deal with the situation when the amount of wind power data is limited, which is a common case in various scenarios. For example, one scenario is the lead time adjustment in online WPF. In this scenario, new forecast models with other specified lead times should be trained, since the uncertainty for the forecast varies with lead times and the performance of the forecast model may deteriorate for forecasting with different forecast horizons. However, considering the real-time forecast efficiency requirement for lead time adjustments, only a limited amount of data is available for training the new forecast model (avoid prolonged training time) leading to overfitting and performance deterioration. Another scenario is WPF for new wind farms. In such a scenario, learning-based forecast algorithms may also suffer from severe overfitting, because there is insufficient historical data (even no historical data). These problems are referred to as \emph{temporal adaptation problem} (for the former scenario) and \emph{spatial adaptation problem} (for the later scenario), respectively.

However, a limitation of traditional deep learning-based approaches is that they rely on a large amount of historical data to train the forecast model via a time-consuming offline learning procedure. In other words, it is difficult for these approaches to deal with situations when 1) high time efficiency is required for training forecast models and 2) the amount of available wind power data is limited. For the situation of high time efficiency requirement in model training, one typical scenario is the lead time adjustment in online WPF. In this scenario, new forecast models with other specified lead times should be trained, since the uncertainty for the forecast varies with lead times and the performance of the forecast model may deteriorate for forecasting with different forecast horizons. However, training another forecast model from scratch with a large amount of data costs a lot of time, which makes it inapplicable for considering the real-time forecast efficiency requirement of lead time adjustments. For the situation of the limited amount of available data, one common case is WPF for new wind farms. Therein, because there is insufficient historical data (even no historical data), learning-based forecast algorithms in this case may suffer from severe overfitting. These problems are referred to as \emph{temporal adaptation problem} (for the former scenario) and \emph{spatial adaptation problem} (for the latter case), respectively.

In contrast to offline learning, online learning extends the model with new observations for considering recent information. This enables adaptivity for the online forecasting strategy by continuously updating forecast algorithms with high time efficiency based on a small amount of online data, which provides possible solutions to above problems facing real-time training efficiency requirements or insufficient data. There are also a few studies concerned about online learning in the WPF literature. For instance, paper \cite{Messner-IJF-2019} proposes a vector autoregression-based WPF model whose parameters are updated recursively by a time-adaptive lasso estimator online. Paper \cite{Kou-AE-2013} formulates an online learning algorithm based on the warped Gaussian process for WPF, which makes the forecast algorithm adaptive to the wind power's characteristics changing with time. Paper \cite{Sommer-IJF-2021} develops an online distributed WPF method based on autoregression to update the forecast model in real-time as well as consider the data privacy issue between different wind farms. Paper \cite{Hu-CSEE-2020} and \cite{Hu-TNLS-2020} design forecast models based on CNN and LSTM, respectively, and update them with incremental learning based on recent observations to automatically consider seasonal and diurnal effects. Paper \cite{Krannichfeldt-TSE-2022} designs an ensemble forecast model whose output is a weighted sum of forecasting results from multiple pre-trained individual forecast models, where weights of individual forecast models' outputs are optimized online with quantile passive aggressive regression. Nevertheless, these online learning approaches need to train offline forecast models with large amounts of historical data first, so they may still encounter severe overfitting due to insufficient historical data when applied to forecasting for newly established wind farms. Moreover, it is still an open question for existing online learning approaches to realize temporal adaptions with different lead times.

%Nevertheless, these online learning approaches need to train offline forecast models with large amounts of historical data first and subsequently update them online only for forecasts with the same lead time determined in the offline procedure. Since the uncertainty of power predictions may be significantly influenced by lead times \cite{Pinson-WE-2007}, these approaches may be not applicable for forecasting with other forecast horizons after online learning based on a few available data. Meanwhile, these methods may also encounter severe overfitting due to insufficient historical data when applied to forecasting for newly established wind farms.

Another possible solution to problems concerning high training efficiency requirements and insufficient data is transfer learning. Transfer learning aims to build a model that possesses a good adaptivity capability in the target domain (with few samples) using knowledge from both the source domain (with sufficient samples) and the target domain \cite{Pan-TKDE-2010}. This means that the model can be directly used in the target domain or applied after fine-tuning based on few samples from the target domain efficiently. The efficacy of transfer learning is mainly endowed by dismissing the difference in data distributions measured by the maximum mean discrepancy between the target domain and the source domain, i.e., finding domain invariant features \cite{Pan-TKDE-2011}. It has been applied to various applications in smart grids, including electric load forecasting \cite{Wu-TSG-2020}, dynamic security assessment \cite{Ren-TPS-2021}, non-intrusive load monitoring \cite{Lin-TSG-2022}, and security-constrained optimal power flow \cite{Liu-TPS-2022}. However, since working conditions of wind farms may change with time and the wind power is nonstationary, the domain invariant features may not be effective for probabilistic WPF under a new condition in the target domain. Thus the performance of the forecast model may deteriorate after transfer learning (also called the negative transfer).

%The above analyses motivate us to study more intelligent algorithms efficiently adapting well to learning tasks with few samples. 
Meta-learning is a prominent few-shot-learning (learning from few samples) approach adapting well to multiple learning tasks \cite{Finn-ICML-2017}. It aims to train a model on a variety of learning tasks, such that it can solve new learning tasks (unseen) in a few-shot-learning manner. In this way, a small number of gradient-update steps with a small amount of training data from new tasks will quickly produce good adaptivity performance on these tasks. The field of meta-learning has seen a dramatic rise in interest in recent years \cite{Hospedales-PAMI-early}. Successful applications of meta-learning have been demonstrated in areas spanning few-shot image classification \cite{Snell-NeurIPS-2017}, unsupervised learning \cite{Metz-ICLR-2019}, reinforcement learning \cite{Alet-ICLR-2020}, and hyperparameter optimization \cite{Franceschi-ICML-2018}. In this context, considering the excellent few-shot-learning ability of meta-learning, it has the potential to create probabilistic WPF algorithms more adaptive to situations where high training efficiency is required and only a limited amount of wind power data is available. %\textcolor{blue}{However, meta-learning ignores the sequential and nonstationary aspects of the probabilistic WPF problems. Unlike image classification problem which has given batches of predetermined labels in a stationary setting, using meta-learning to develop approaches adapting to the sequential and nonstationary wind power data is challenging.}

In this paper, we introduce meta-learning to the probabilistic WPF to enhance the performance of forecasts with different lead times online where model training requires high time efficiency and forecasts for the newly established wind farms where the amount of historical wind power data is limited. The contributions of this paper are as follows:

%\hangafter=1 \hangindent 2.1em 1) A ``task stream'' is proposed to evaluate the temporal and spatial adaptabilities of forecast models. Specifically, we first formulate different forecast tasks in the probabilistic WPF based on temporal and spatial information, i.e., forecasting with different lead times and location information. Then, we define the ``task stream'' for online forecasting to evaluate the forecast model's performance over different online forecast tasks in a continual few-shot-learning setting. 

\hangafter=1 \hangindent 2em 1) A two-part learning approach is designed based on \emph{meta-learning} for probabilistic WPF. The proposed method includes offline and online learning procedures. In the offline learning part, a base forecast model is trained via inner and outer loop updates of meta-learning to equip it with temporal and spatial adaptabilities. Then, in the online learning part, the trained base forecast model quickly adapts to different forecast tasks (forecasting with different lead times or locations) in a continual few-shot-learning setting based on the newly collected wind power data online. To the best of our knowledge, this is the first study to apply meta-learning to probabilistic WPF problems.

%A two-part learning approach, including offline learning and online learning, is designed based on \emph{meta-learning} \cite{Finn-ICML-2017} for probabilistic WPF combined with online learning and quantile regression \cite{Taieb-TSG-2016}. To the best of our knowledge, this is the first time that meta-learning has been applied to probabilistic WPF. Specifically, in the offline learning part, a base forecast model is trained via inner and outer loop updates of meta-learning to equip it with great temporal and spatial adaptabilities. Then, in the online learning part, the trained base forecast model quickly generalizes well on different forecast tasks (forecasting with different lead times or locations) in a continual few-shot-learning setting. Meanwhile, the nonstationary assumption of data distributions is intrinsically held here since the forecast model continuously adapts to the newly collected wind power data online. 

\hangafter=1 \hangindent 2em 2) Two applications have been developed based on our proposed two-part learning approach: temporal adaptations about probabilistic WPF with different lead times, and spatial adaptations about probabilistic WPF for the newly established wind farms. The temporal and spatial adaptabilities of corresponding developed methods for applications were corroborated under a comprehensive verification framework, considering the accordance with reality, prediction interval (PI) width as well as the accuracy of 0.5-$th$ quantiles (median).
%Two applications concerning probabilistic WPF with different lead times and probabilistic WPF for newly established wind farms have been developed based on our proposed two-part learning approach, respectively. The temporal and spatial adaptabilities of corresponding developed methods for applications were corroborated under a comprehensive verification framework considering the accordance with reality, prediction interval (PI) width as well as the accuracy of 0.5-$th$ quantiles (median).

The remainder of this article is organized as follows. Section \uppercase\expandafter{\romannumeral2} presents preliminaries about meta-learning, the basic model for probabilistic WPF, and evaluation metrics. Section \uppercase\expandafter{\romannumeral3} introduces the definition of forecast tasks and the proposed approach for the probabilistic WPF. Section \uppercase\expandafter{\romannumeral4} and  Section \uppercase\expandafter{\romannumeral5} give applications considering temporal and spatial adaptations in the probabilistic WPF based on our proposed approach, respectively. Section \uppercase\expandafter{\romannumeral6} draws conclusions.

% === II. PRELIMINARY =============================================================
% =================================================================================
\vspace{-0.2cm}
\section{Preliminary}
\vspace{-0.09cm}
\subsection{Meta-learning}
\begin{figure}[h]
  \begin{center}
  \includegraphics[width=2.4in]{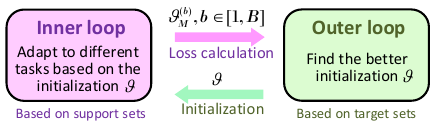}\\
  \vspace{-0.5cm}
  \caption{The inner loop and outer loop in meta-learning.}\label{inner-outer}
  \end{center}
  \vspace{-0.7cm}
\end{figure}

The meta-learning algorithm \cite{Finn-ICML-2017} finds an initialization (also called meta-parameters) for a neural network (NN) so that new tasks can be learned quickly through few-shot learning (learning from few samples). More formally, we define a NN as a \emph{base model} whose parameters are used as meta-parameters for upcoming learning tasks. Representing the base model as $f_{\vartheta}$, where $\vartheta$ denote meta-parameters, the best $\vartheta$ are determined via training the base model to adapt to multiple tasks with two optimization loops, i.e., inner and outer loops, which are shown in Fig.~\ref{inner-outer}. 
\vspace{-0.5cm}

In the inner loop, we adapt the base model with meta-parameters $\vartheta$ to different learning tasks. Specifically, different data sets, i.e., support sets, are sampled for different tasks first, which are used for model training in the inner loop (inner loop updates). Given a task $b$, let $\vartheta_{m}^{(b)}$ signify $\vartheta$ after $m$ gradient updates via gradient descent. During each update, we compute
\vspace{-0.4cm}
\begin{equation}
  \vartheta_m^{(b)}=\vartheta_{m-1}^{(b)}-\alpha_{in}\nabla_{\vartheta_{m-1}^{(b)}}\mathcal{L}_{S^{(b)}}(f_{\vartheta_{m-1}^{(b)}(\vartheta)})\text{, }
  \label{inner-update}
  \vspace{-0.2cm}
\end{equation}
for $m$ fixed across all tasks, where $\alpha_{in}$ is the inner loop learning rate, $S^{(b)}$ is the support set for task $b$, and $\mathcal{L}_{S^{(b)}}(f_{\vartheta_{m-1}^{(b)}(\vartheta)})$ is the loss on the support set of task $b$ after $m-1$ updates, making clear the dependence of $f_{\theta_{m-1}^{(b)}}$ on $\vartheta$. This set of multiple update steps is called the inner-loop update process. The total number of update steps for the inner-loop update process is set as a small value since we want the base model to adapt to different tasks via a small number of gradient-update steps and the training process is efficient.

In the outer loop, we update meta-parameters $\vartheta$ based on updating results in  (\ref{inner-update}) for better meta-parameters initializing in the inner loop. Specifically, additional data sets, i.e., target sets, are sampled for different tasks for model training in the outer loop (outer loop updates). A meta-loss is defined based on target sets as
\vspace{-0.3cm}
\begin{equation}
  \mathcal{L}_{meta}(\vartheta)=\sum_{b=1}^B\mathcal{L}_{T^{(b)}}(f_{\theta_{M}^{(b)}(\vartheta)})\text{, }
  \label{meta-loss}
  \vspace{-0.2cm}
\end{equation}
where $M$ is the total number of inner loop updates, $B$ is the number of tasks, $T^{(b)}$ is the target set for task $b$, and $\mathcal{L}_{T^{(b)}}(f_{\theta_{M}^{(b)}(\vartheta)})$ is the loss on the target set of task $b$ after $M$ inner loop updates. The loss (\ref{meta-loss}) measures the quality of an initialization $\vartheta$ in terms of the total loss of using that initialization across all tasks. This meta-loss is now minimized to optimize the initial parameter value $\vartheta$ considering the across-task knowledge. The optimization of this meta-loss is called the outer-loop update process. 

The resulting update for the meta-parameters $\vartheta$ with an outer loop learning rate $\alpha_{out}$ can be expressed as
\vspace{-0.2cm}
\begin{equation}
  \vartheta = \vartheta-\alpha_{out}\nabla_{\vartheta}\mathcal{L}_{meta}(\vartheta)
  \label{update-meta}
  \vspace{-0.2cm}
\end{equation}

In summary, the inner loop takes meta-parameters from the outer loop, and separately performs a few gradient updates for each task over the support set provided for adaptation. The outer loop updates meta-parameters over the target set to a setting that enables fast adaptation to different tasks. The iterative implementation of outer loop and inner loop following (\ref{inner-update}) to (\ref{update-meta}) will not stop until meets the maximum training epoch. In this way, we obtain the base model with the ability to fast adapt to new tasks.

\vspace{-0.4cm}
\subsection{Basic Model for Probabilistic WPF}
\vspace{-0.1cm}
Distribution types of wind power (view it as a random variable) may change with time influenced by capricious weather conditions. Nonparametric methods thus are more adaptive than parametric ones as no pre-defined distributional assumptions are required. In this context, we use quantile regression, a widely used nonparametric approach, for the probabilistic WPF \cite{Taieb-TSG-2016}. The basic model for probabilistic WPF can be expressed as
\vspace{-0.3cm}
\begin{equation}
  {\hat y}_{t+\tau|t}^{q}=F(\boldsymbol X_t\text{, }q\text{; }\theta)\text{, }q\in Q\text{, }
  \label{quantile-model}
  \vspace{-0.2cm}
  \end{equation}
where ${\hat y}_{t+\tau|t}^{q}$ is the $q$-$th$ quantile prediction of wind power with the lead time $\tau$, $F(\cdot\text{; }\theta)$ is the probabilistic forecast model parameterized by $\theta$, $Q$ is a quantile set including all quantiles concerned, and $\boldsymbol X_t$ is a vector concatenated with different available time-series features covering a lag interval $\delta$ from time $t-\delta+1$ to time $t$. The probabilistic forecast model is trained via minimizing the pinball loss $\mathcal{L}$, which is formulated as
\vspace{-0.2cm}
\begin{equation}
  % \resizebox{0.9\hsize}{!}{
  \mathcal{L}=\sum_{q\in Q}q\cdot\text{max}(0, y_{t+\tau}-\hat{y}_{t+\tau|t}^{q})+(1-q)\cdot\text{max}(0,\hat{y}_{t+\tau|t}^{q}-y_{t+\tau})\text{,}
  \label{pinball-loss}
  \vspace{-0.1cm}
  \end{equation}
where $y_{t+\tau}$ is the observation of $\hat y_{t+\tau|t}^{q}$ (also named as the forecasting target). This loss function aims to provide a forecast with a $q$ probability of under forecasting the observation and a ($q-1$) probability of over forecasting the observation. 

\vspace{-0.5cm}
  \subsection {Evaluation Metrics}
  The forecasted quantiles are evaluated within the probabilistic forecast evaluation framework in \cite{Pinson-WE-2007}. Besides, the forecasted 0.5-$th$ quantile is taken as an improved point forecasting result and evaluated with the statistical accuracy measurement in \cite{Wang-TPS-2018}.
  
  \subsubsection {Reliability}
  The overall reliability of a probabilistic forecast model measures average deviations between the nominal proportion and the observed frequency of the data below the quantile forecasting as
  \vspace{-0.2cm}
  \begin{equation}
  \overline{b_\tau}=\frac{1}{J}\sum_{j=1}^J|q_{j}-\frac{1}{N}\sum_{i=1}^NH(\hat{y}_{i+\tau|i}^{q_j}-y_{i+\tau})|\text{,}
  \label{b_tau}
  \vspace{-0.2cm}
  \end{equation}
  where $N$ is the number of samples in the testing data set, $q_{j}$ is the nominal proportion from $q_{1}=5\%$ to $q_{J}=95\%$ ($J=19$) with steps 5\%, and $H(x)$ is the unit step function.% whose value is 1 if $x\geq0$, otherwise 0. Variables $\hat{y}_{i+\tau|i}^{q_j}$ (or $y_{i+\tau}$) with different subscripts can belong to different forecast tasks and we do not distinguish them here for brevity.
  % \begin{equation}
  % H(x)=\left\{
  % \begin{array}{l}
  % 1\text{, }x\geq0\\
  % 0\text{. }x<0
  % \end{array}\text{.}
  % \right.
  % \end{equation}
  
  \subsubsection {Sharpness}
  Sharpness measures the average width of PIs with different (1-$q_{j}$) as
  \vspace{-0.2cm}
  \begin{equation}
  {\overline{\delta_\tau}}=\frac{1}{J\cdot N}\sum_{j=1}^J\sum_{i=1}^N(\hat{y}_{i+\tau|i}^{1-q_{j}/2}-\hat{y}_{i+\tau|i}^{q_{j}/2})\text{.}
  \label{d_tau}
  \vspace{-0.2cm}
  \end{equation}
  
  \subsubsection {Skill Score}
  Skill score takes both reliability and sharpness into consideration and an average skill score for $N$ time spots is defined as
  \vspace{-0.2cm}
  \begin{equation}
  \overline{S_\tau}=\frac{1}{N}\sum_{i=1}^N\sum_{j=1}^J \{[H(\hat{y}_{i+\tau|i}^{q_{j}}-y_{i+\tau})-q_{j}](y_{i+\tau}-\hat{y}_{i+\tau|i}^{q_{j}})\}\text{.}
  \label{s_tau}
  \vspace{-0.2cm}
  \end{equation}
  
  \subsubsection {Mean Absolute Error}
  Mean absolute error (MAE) measures absolute deviations between the point forecasting result and the observation as
  \vspace{-0.2cm}
  \begin{equation}
    MAE=\frac{1}{N}\sum_{i=1}^N |y_{i+\tau}-\hat{y}_{i+\tau|i}^{0.5}|.
    \label{mae}
    \vspace{-0.2cm}
  \end{equation}

\section{Probabilistic WPF With Meta-Learning}

In this section, the meta-learning-based approach for probabilistic WPF is presented. We first introduce the forecast task defined in this paper. Subsequently, an offline learning part and an online learning part associated with different forecast tasks are detailed for the probabilistic WPF.
\vspace{-0.4cm}
\subsection {Forecast Task}

In this paper, the forecast of wind power outputs of wind farm $l$ with lead time $\tau$ is referred to as a forecast task, and different $l$ or $\tau$ specifies different forecast tasks. We assume all forecast tasks are drawn from a same distribution of tasks. For each forecast task $n$, we have a loss function $\mathcal{L}_n$ and a sampled data set $\mathbb{D}^{(n)}$. Specifically, the loss function $\mathcal{L}_n$ is the pinball loss defined in (\ref{pinball-loss}) and the data set $\mathbb{D}^{(n)}=\{(\boldsymbol X_t\text{, }y_{t+\tau})\}$ is generated with historical observations according to the lead time and location. To apply the forecast model to different forecast tasks, here we set the size of $\boldsymbol X_t$ and $y_{t+\tau}$ for different forecast tasks to be equal, respectively. Denote a probabilistic forecast model as $F(\cdot\text{; }\theta)$ parameterized by $\theta$ as presented in (\ref{quantile-model}). We update $\theta$ to adapt $F(\cdot\text{; }\theta)$ to forecast task $n$ via minimizing a loss function $\mathcal{L}_{\mathbb{D}^{(n)}}$ based on the data set $\mathbb{D}^{(n)}$.
\vspace{-0.4cm}
\subsection {The Offline Learning Part}

\begin{figure}[h]
  \begin{center}
  \includegraphics[width=3.5in]{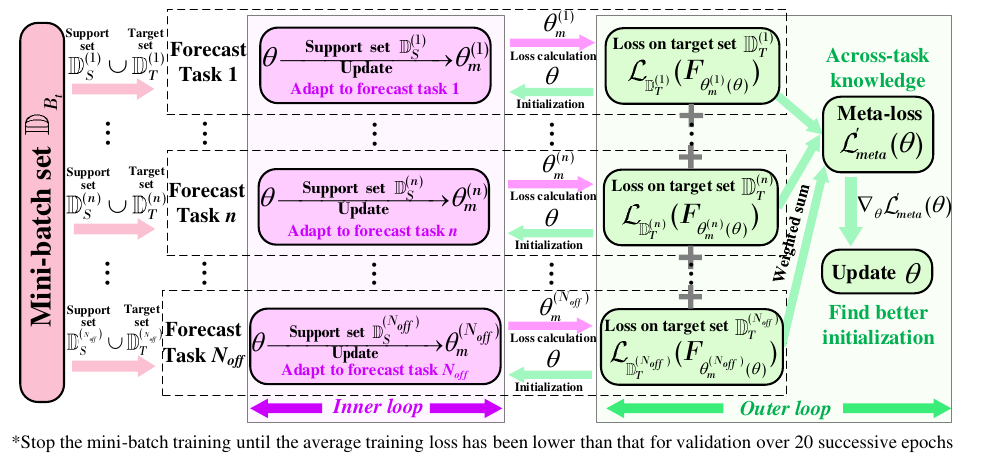}\\
  \vspace{-0.3cm}
  \caption{The offline learning part for probabilistic WPF.}\label{Offline part}
  \end{center}
  \vspace{-0.5cm}
\end{figure}

In the offline learning part, we apply inner and outer loop updates of meta-learning to the training of a \emph{base forecast model}, which equips the base forecast model with adaptability to different forecast tasks. Denote the base forecast model as $F(\cdot\text{; }\theta)$, where $\theta$ represents \emph{meta-parameters} here. The aim of the offline learning part is to find the best meta-parameters of the base forecast model by adapting the model to different forecast tasks. Before the model training, mini-batch sets of data should be sampled. Assuming there are $N_{off}$ offline forecast tasks, we generate data sets $\{\mathbb{D}^{(n)}\}_{n=1}^{N_{off}}$ for all forecast tasks according to historical observations. A mini-batch set $\mathbb{D}_{B_{t}}$ is randomly sampled from the joint data set $\mathbb{D}^{(1)}\cup\cdots\mathbb{D}^{(n)}\cdots\cup\mathbb{D}^{(N_{off})}$ for the mini-batch training, where $B_{t}$ is the number of samples in $\mathbb{D}_{B_{t}}$. The base forecast model is trained with mini-batch sets via inner and outer loops as visualized in Fig.~\ref{Offline part}, which is detailed as below. 
\vspace{-0.2cm}

First, the mini-batch set $\mathbb{D}_{B_{t}}$ is divided into sub-sets $\{\mathbb{D}_S^{(n)}\cup\mathbb{D}_T^{(n)}\}_{n=1}^{N_{off}}$ for different forecast tasks during each training step, where $\mathbb{D}_S^{(n)}$ is the support set used in the inner loop for forecast task $n$, and $\mathbb{D}_T^{(n)}$ is the target set used in the outer loop for the same forecast task, i.e., forecast task $n$ here.

% \begin{algorithm}
%   \caption{Sampling Technique for Forecast Tasks}\label{algorithm 1}
%   \KwIn{Historical observations, total task number $N_{off}$, batch size $B_{task}$, $b=1$}
%   \KwOut{Mini-batches of data for forecast tasks}
%   Generate data sets $\mathbb{D}_1\text{, }\cdots\mathbb{D}_n\text{, }\cdots\mathbb{D}_{N_{off}}$, where $cat(n\text{, }\boldsymbol X_t\text{, }y_{t+\tau})\in\mathbb{D}_n$\;
%   \While{$n\leq N_{off}$}
%   {Randomly sample a subset $\mathbb{D}_{B_{task}}$ from the joint set $\mathbb{D}_1\cup\cdots\mathbb{D}_n\cup\cdots\mathbb{D}_{N_{off}}$ with batch size $B_{task}$\;
%   Obtain $\mathbb{D}_{B_{task}}^{(n)}$ for the forecast task $n$ according to the label information in $\mathbb{D}_{B_{task}}$\;
%   \If{$\mathbb{D}_{B_{task}}^{(n)}$ is not $\varnothing$}{Record $\mathbb{D}_{B_{task}}^{(n)}$;}
%   $n=n+1$\;}
% \end{algorithm}

Second, in the inner loop, we adapt the base forecast model to different forecast tasks. Let $\theta_{m}^{(n)}$ signify parameters after $m$ gradient updates for the forecast task $n$ via gradient descent, we compute
\vspace{-0.2cm}
\begin{equation}
  \theta_m^{(n)}=\theta_{m-1}^{(n)}-\alpha_{in}\nabla_{\theta_{m-1}^{(n)}}\mathcal{L}_{\mathbb{D}_{S}^{(n)}}(F_{\theta_{m-1}^{(n)}(\theta)})\text{, }
  \label{r-inner-loop}
  \vspace{-0.2cm}
\end{equation}
where $F_{\theta_{m-1}^{(n)}(\theta)}$ denotes the base forecast model $F(\cdot\text{; }\theta)$ after $m-1$ updates, and $\mathcal{L}_{\mathbb{D}_{S}^{(n)}}(F_{\theta_{m-1}^{(n)}(\theta)})$ denotes the corresponding loss on $\mathbb{D}_{S}^{(n)}$ of forecast task $n$. 

Third, in the outer loop, meta-parameters $\theta$ are updated based on results in (\ref{r-inner-loop}). In the basic meta-learning framework, only the loss of the final step in the inner loop process contributes to the meta-loss. However, this may cause instability in the training process since the model's parameters at every step except the last one are optimized implicitly as a result of backpropagation \cite{meta train}. Therefore, here we propose to consider the loss calculated in each inner loop update and reformulate the meta-loss in (\ref{meta-loss}) as a weighted sum
\vspace{-0.2cm}
\begin{equation}
  \mathcal{L}^{\prime}_{meta}(\theta)=\sum_{n=1}^{N_{off}}\sum_{m=1}^M\frac{m}{M}\mathcal{L}_{\mathbb{D}_{T}^{(n)}}(F_{\theta_{m}^{(n)}(\theta)})\text{, }
  \label{r-meta-loss}
  \vspace{-0.1cm}
\end{equation}
where $\frac{m}{M}$ denotes a decay rate decreasing the contributions from earlier steps and slowly increasing the contribution of later steps in inner loop updates. This is done to ensure that as training progresses the final step loss receives more attention from the optimizer thus ensuring it reaches the lowest possible loss. It should be noticed that if either $\mathbb{D}_{S}^{(n)}$ or $\mathbb{D}_{T}^{(n)}$ is $\varnothing$, we skip to (\ref{r-inner-loop}) and (\ref{r-meta-loss}) for the next forecast task, i.e., the forecast task $n+1$. Subsequently, meta-parameters are updated as 
\vspace{-0.2cm}
\begin{equation}
\theta = \theta-\alpha_{out}\nabla_{\theta}\mathcal{L}^{\prime}_{meta}(\theta)\text{,}
\label{orig-grad}
\vspace{-0.2cm}
\end{equation}
and we reinitialize the inner loop with $\theta$. 

Next, we introduce an approximation approach to accelerate the training process. Considering the gradient update step of $\theta_i$ (the $ith$ entry in $\theta$) in (\ref{orig-grad}), the gradient of $\mathcal{L}^{\prime}_{meta}(\theta)$ with respect to $\theta_i$ can be formed as
\vspace{-0.2cm}
\begin{equation}
  \begin{aligned}
    &\nabla_{\theta_{i}}\mathcal{L}^{\prime}_{meta}(\theta)=\sum_{n=1}^{N_{off}}\sum_{m=1}^M\frac{m}{M} \sum_j\frac{\partial\mathcal{L}_{\mathbb{D}_{T}^{(n)}}(F_{\theta_{m}^{(n)}(\theta)})}{\partial\theta_{m_j}^{(n)}(\theta)} \frac{\partial\theta_{m_j}^{(n)}(\theta)}{\partial\theta_i}.
    \label{grad-metaloss-theta} 
  \end{aligned}
  \vspace{-0.1cm}
  \end{equation}

In (\ref{grad-metaloss-theta}), the computing of $(\partial\theta_{m_j}^{(n)}(\theta))/ (\partial\theta_i)$ relies on past updates of $\theta$. We first expand $\theta_{m_j}^{(n)}(\theta)$ as
\vspace{-0.2cm}
\begin{equation}
  \theta_{m_j}^{(n)}(\theta)=\theta_j-\alpha_{in}\sum_{k=0}^{m-1}\frac{\partial\mathcal{L}_{\mathbb{D}_{T}^{(n)}}(F_{\theta_{k}^{(n)}(\theta)})}{\partial\theta_{k_j}^{(n)}} 
  \vspace{-0.2cm}
\end{equation}

\vspace{-0.1cm}
Then, $(\partial\theta_{m_j}^{(n)}(\theta))/ (\partial\theta_i)$ can be calculated as
\vspace{-0.2cm} 
\begin{align}
  \frac{\partial\theta_{m_j}^{(n)}(\theta)}{\partial\theta_i} =\left\{
  \begin{array}{l}
  1-\rho\text{ , if }i=j\text{, }  \\
  -\rho\text{ , if }i\neq j.
  \end{array}
  \right.\text{,}
  \label{grad-2nd}
  \vspace{-0.5cm} 
\end{align}
where $\rho$ denotes $\alpha_{in}\sum_{k=0}^{m-1}\frac{\partial\mathcal{L}_{\mathbb{D}_{T}^{(n)}}^2(F_{\theta_{k}^{(n)}(\theta)})}{\partial\theta_{k_j}^{(n)}\partial\theta_i}$. We see that $\rho$ involves second-order gradients, thus it is very time-consuming. To reduce the computation burden in (\ref{orig-grad}), we adopt a gradient approximation method mentioned in \cite{Finn-ICML-2017}. Considering that the learning rate $\alpha_{in}$ is close to zero and the gradient propagating through the NN is generally a finite value and will not be extremely large, $\rho$ will be close to zero as well. Therefore, (\ref{grad-metaloss-theta}) can be approximated as
\vspace{-0.2cm}
\begin{equation}
  \nabla_{\theta_{i}}\mathcal{L}^{\prime}_{meta}(\theta)\approx\sum_{n=1}^{N_{off}}\sum_{m=1}^M\frac{m}{M} \frac{\partial\mathcal{L}_{\mathbb{D}_{T}^{(n)}}(F_{\theta_{m}^{(n)}(\theta)})}{\partial\theta_{m_i}^{(n)}(\theta)}\text{. } 
  \label{grad-approx} 
  \vspace{-0.2cm}
\end{equation}
In (\ref{grad-approx}) only the first-order derivative is needed. 

Nevertheless, the above gradient approximation through the whole training process may reduce the adaptivity of the base forecast model to different forecast tasks. We thus implement the gradient approximation in (\ref{grad-approx}) first until the training loss is lower than a certain threshold, and then switch to the second-order gradient in (\ref{grad-metaloss-theta}) for the remainder of the training phase. Gradient updates of other parameters in $\theta$ can be dealt with the same as $\theta_i$ and we redo the update in (\ref{orig-grad}). Intuitively, using the gradient approximation before starting to use second-order gradients can be viewed as a strong pre-training method that quickly offers a warm start for the offline learning part.

In summary, during each iteration in the offline learning part, we first sample mini-batches of data for forecast tasks. Then, based on sampling results, we implement inner loop and outer loop updates to update meta-parameters of the base forecast model. A weighted sum-form meta-loss and a gradient approximation approach are introduced in training the base forecast model to stabilize and speed up the training process as suggested in \cite{meta train,Finn-ICML-2017}. We iterate this process on all sampled mini-batches and stop it until the average training loss has been lower than that for validation over 20 successive epochs.

% This offline learning part equips the base forecast model with a prominent ability to adapt to different forecast tasks, which is useful for the temporal adaptation () and spatial adaptation () in probabilistic WPF.
\vspace{-0.5cm}
\subsection {The Online Learning Part}
\vspace{-0.1cm}

\begin{figure}[h]
  \begin{center}
  \includegraphics[width=3.2in]{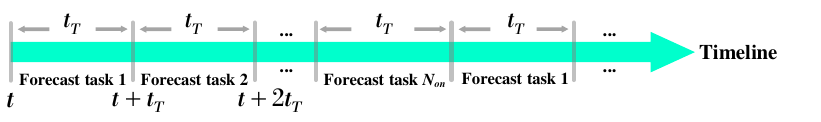}\\
  \vspace{-0.5cm}
  \caption{Illustration for the ``task stream''.}\label{task stream}
  \end{center}
  \vspace{-0.6cm}
\end{figure}

In the online learning part, we adapt the forecast model to different online forecast tasks. To evaluate the adaptivity of the forecast model for different forecast tasks, we put different forecast tasks along the timeline and formulate a ``task stream'' as shown in Fig.~\ref{task stream}. Specifically, assume that we have $N_{on}$ kinds of online forecast tasks in the online learning part and each forecast task lasts for a same period $t_{T}$ for simplicity. We implement forecast tasks from 1 to $N_{on}$ on the whole data set for the adaptivity capability evaluation of the forecast model. By changing $t_{T}$, we can set different switching frequencies for online forecast tasks, which imitate different forecasting requirements in a real-world setting.
\vspace{-0.4cm}
\vspace{-0.2cm}
\begin{figure}[h]
  \begin{center}
    \vspace{-0.2cm}
  \includegraphics[width=3.2in]{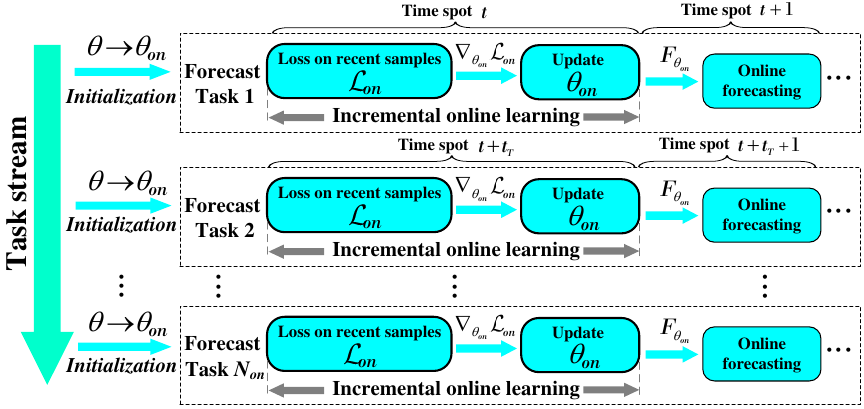}\\
  \vspace{-0.3cm}
  \caption{The online learning part for probabilistic WPF.}\label{Online part}
  \end{center}
  \vspace{-0.6cm}
\end{figure}

Then, the online learning of forecast models is implemented incrementally to adapt tasks in the ``task stream'', which is presented in Fig.~\ref{Online part}. In detail, when we start the online application or every time change the forecast task in the ``task stream'' (after $t_{T}$), the forecast model will be initialized/reinitialized with meta-parameters $\theta$ learned in the offline learning part. This utilizes the adaptivity of the base forecast model to different forecast tasks. Subsequently, we formulate an online loss $\mathcal{L}_{on}$ for the forecast task with recent $N_{\lambda}$ samples and a forgetting factor $\lambda\in[0\text{, }1]$, which is defined as
\vspace{-0.2cm}
\begin{equation}
  \mathcal{L}_{on}=\frac{1}{N_{\lambda}}\sum_{i=t-\tau-N_{\lambda}+1}^{t-\tau} \lambda^{t-\tau-i}\cdot \mathcal{L}_{\{(\boldsymbol X_i\text{, }y_{i+\tau})\} } (F_{\theta_{on}})\text{, } 
  \label{loss-on} 
  \vspace{-0.2cm}
\end{equation}
where $\theta_{on}$ are meta-parameters $\theta$ when we start the online application or change the forecast task in the ``task stream'' and $F_{\theta_{on}}$ is the forecast model online. The $\mathcal{L}_{\{(\boldsymbol X_i\text{, }y_{i+\tau})\} }(F_{\theta_{on}})$ is the loss calculated from the forecast model by taking $(\boldsymbol X_i\text{, }y_{i+\tau})\text{, }i\in[t-\tau-N_{\lambda}+1\text{, }t-\tau]$ as inputs, and we omit the superscript label of the forecast task for conciseness. Next, parameters of $F_{\theta_{on}}$ are updated with an online learning rate $\alpha_{on}$ 
\vspace{-0.2cm}
\begin{equation}
  \theta_{on} = \theta_{on}-\alpha_{on}\nabla_{\theta_{on}}\mathcal{L}_{on}.
  \label{update-on}
  \vspace{-0.2cm}
\end{equation}

Eq. (\ref{loss-on}) and (\ref{update-on}) update the forecast model $F_{\theta_{on}}$ incrementally with the newest observations and we name them as the \emph{incremental online learning} with a rolling-window manner. Incremental online learning may iterate several times, denoted by $N_{inc}$, for the same samples in the sliding window (during the time gap waiting for data at the next time spot $t+1$) for higher performance on online forecast tasks. Subsequently, after collecting the input $\boldsymbol X_{t+1}$ at time spot $t+1$, the trained forecast model takes in $\boldsymbol X_{t+1}$ and generates online forecasting results at time spot $t+\tau+1$. The incremental online learning and online forecasting are implemented alternately on the whole data set of online forecast tasks in the ``task stream''.

The proposed two-part learning method only needs to train a based forecast model offline and apply it to online applications with fast online adaptation. This avoids pre-training for a large number of models (e.g., using a transfer learning technique \cite{Liu-TPS-2022}), which decreases computational cost and storage complexity especially when we want to implement forecast tasks with different lead times related to many wind farms. In the next two sections, based on the proposed meta-learning-based approach for probabilistic WPF, we develop two applications considering forecasting with different lead times and forecasting for the newly established wind farms, respectively.
\vspace{-0.3cm}
\section {Application \uppercase\expandafter{\romannumeral1}: Forecasting With Different Lead Times}
\vspace{-0.1cm}
\subsection {Problem Description}
Probabilistic forecast models are generally designed for a specified lead time. However, the uncertainty of WPF varies with forecast horizons \cite{Pinson-WE-2007}, which impedes the adaptivity between probabilistic forecast models with different lead times, especially when only a small amount of data is used for model training/updating to meet real-time requirements in online applications. This makes the fast online adjustment of lead times in online forecasting a very challenging task.
% \textcolor{blue}{Meanwhile, it may also not be feasible to maintain a set of pre-trained models with different pre-determined lead times and directly apply them to the corresponding forecast tasks online, as the actual needs online for forecast horizons may be unpredictable. These problems make the fast online adjustment of lead times in online forecasting a very challenging task.}

To improve the adaptivity between forecast models with different forecast horizons and make fast online adjustments of lead times possible, we propose to formulate a base forecast model considering different uncertainty information of forecast horizons and make it quickly adapt well to forecasts with different lead times. This application demonstrates the adaptivity between different lead times of forecast models, i.e., the temporal adaptation.
\vspace{-0.4cm}
\subsection {Application Method}
The detailed method for this application can be realized by the two-part learning approach proposed in section \uppercase\expandafter{\romannumeral3}-B and \uppercase\expandafter{\romannumeral3}-C, where we refer to forecasts of wind power outputs with different lead times as different forecast tasks. Specifically, in the offline learning part, the base forecast model is trained with mini-batch sets sampled from all generated data sets for different forecast tasks via inner and outer loops visualized in Fig.~\ref{Offline part}. In the inner loop, we adapt the base forecast model to forecast tasks associated with different lead times via (\ref{r-inner-loop}). In the outer loop, a meta-loss evaluating the adaptability of the base forecast model to all forecast tasks with different lead times is calculated based on the updating results in the inner loop via (\ref{r-meta-loss}), and the base forecast model is updated via (\ref{orig-grad}). In (\ref{r-meta-loss}), the loss calculated from different forecast tasks shares the same coefficient, since we assume a uniform distribution over forecast tasks with different lead times that we want the base forecast model to be able to adapt to. We iterate inner and outer loop updates on all sampled mini-batch sets and stop them with the early stopping strategy presented at the end of section \uppercase\expandafter{\romannumeral3}-B. In this way, we obtain the base forecast model with optimized meta-parameters $\theta$ and endow the model with across-task knowledge of different lead times.

In the online learning part, we first formulate a ``task stream'' with many online forecast tasks associated with different lead times (different from those in the offline learning part) as demonstrated in Fig.~\ref{task stream}. Then, following the information flow presented in Fig.~\ref{Online part}, the trained base forecast model is used in the ``task stream'' for online forecasting combined with incremental online learning via (\ref{loss-on}) and (\ref{update-on}) to adapt to tasks with different forecast horizons. To clearly illustrate how to use the trained base forecast model and make forecasts with different lead times online, we show the relationship between probabilistic forecast models for online forecasting at two successive time spots $t$ and $t+1$ under two conditions:

\subsubsection {Forecasting With the Same Lead Time $\tau$ at Time Spots $t$ and $t+1$}
An online loss $\mathcal{L}_{t}$ at time spot $t$ is calculated following (\ref{loss-on}) as $\mathcal{L}_{t}=\frac{1}{N_{\lambda}}\sum_{i=t-\tau-N_{\lambda}+1}^{t-\tau} \lambda^{t-\tau-i}\cdot \mathcal{L}_{\{(\boldsymbol X_i\text{, }y_{i+\tau})\} } (F_{\theta_{t}})$,
% \vspace{-0.2cm}
% \begin{equation}
%   \mathcal{L}_{t}=\frac{1}{N_{\lambda}}\sum_{i=t-N_{\lambda}+1}^t \lambda^{t-i}\cdot \mathcal{L}_{\{(\boldsymbol X_i\text{, }y_{i+\tau})\} } (F_{\theta_{t}})\text{, } 
%   \label{loss-on-lt1}
% \vspace{-0.2cm}
% \end{equation}
where $F_{\theta_{t}}$ is the forecast model for online forecasting at time spot $t$ parameterized by $\theta_{t}$. %Particularly, $\theta_{t}$ are the optimized meta-parameters $\theta$ if the forecast task begins with time spot $t$. 
Then, parameters for $F_{\theta_{t+1}}$ at time spot $t+1$ are calculated as $\theta_{t+1} = \theta_{t}-\alpha_{on}\nabla_{\theta_{t}}\mathcal{L}_{t}.$ Particularly, if the online application starts with time spot $t$, a loss $\mathcal{L}_{st}$ will be formed based on the optimized meta-parameters $\theta$ as $\mathcal{L}_{st}=\frac{1}{N_{\lambda}}\sum_{i=t-\tau-N_{\lambda}+1}^{t-\tau} \lambda^{t-\tau-i}\cdot \mathcal{L}_{\{(\boldsymbol X_i\text{, }y_{i+\tau})\} } (F_{\theta})$ first, then parameters of $F_{\theta_{t}}$ are calculated as  $\theta_{t} = \theta-\alpha_{on}\nabla_{\theta}\mathcal{L}_{st}.$ 
% \vspace{-0.2cm}
% \begin{equation}
%   \theta_{t+1} = \theta_{t}-\alpha_{t}\nabla_{\theta_{t}}\mathcal{L}_{t}.
%   \label{update-on-lt1}
%   \vspace{-0.2cm}
% \end{equation}

\subsubsection {Forecasting With Different Lead Times $\tau_1$ and $\tau_2$ at Time Spots $t$ and $t+1$, Respectively}
Forecast model for online forecasting at time spot $t$ is not used at time spot $t+1$. Instead, an online loss $\mathcal{L}_{t}$ at time spot $t$ is calculated based on the optimized meta-parameters $\theta$ as $\mathcal{L}_{t}=\frac{1}{N_{\lambda}}\sum_{i=t-\tau_2-N_{\lambda}+1}^{t-\tau_2} \lambda^{t-\tau_2-i}\cdot \mathcal{L}_{\{(\boldsymbol X_i\text{, }y_{i+\tau_2})\} } (F_{\theta})$. Then, parameters for $F_{\theta_{t+1}}$ at time spot $t+1$ are calculated as $\theta_{t+1} = \theta-\alpha_{on}\nabla_{\theta}\mathcal{L}_{t}.$

In this way, we make full use of the knowledge accumulated in the same forecast task and the adaptability of the base forecast model to the forecast with different lead times. 

\vspace{-0.4cm}
\subsection {Setting of Numerical Simulations}
\subsubsection{Benchmarks}
We consider representative works in \cite{Hu-TNLS-2020} and \cite{Finn-ICML-2017} in our comparisons, which are described as follows.

$\bullet$ Single-task learning, where the probabilistic forecast model was trained based on sufficient historical data corresponding to one single forecast task \cite{Hu-TNLS-2020}. Here, the trained forecast model was evaluated with the same online learning part with the ``task stream'' proposed in the subsection \uppercase\expandafter{\romannumeral3}-C. To achieve the best online performance, we trained the probabilistic forecast model for each offline forecast task and chose the one with the best performance in the validation phase for the evaluation and comparison in the online learning part. 

$\bullet$ Multi-task learning through averaging the output space (MTAO), which is an approach adapting to learning tasks with few samples \cite{Finn-ICML-2017}. The detailed implementation of MTAO can be found on page 5 of \cite{Finn-ICML-2017}. Based on this setting, we trained a probabilistic forecast model with data corresponding to all offline forecast tasks during the offline learning part. This can be viewed as pre-training on all tasks, so we directly integrated gradients from all tasks and update parameters of the probabilistic forecast model. As a result, the probabilistic forecast model learned the average output among forecast tasks with the same input. The online learning part of MTAO was the same as our proposal.

$\bullet$ Multi-task learning through averaging the parameter space (MTAP), which is another approach adapting to learning tasks with few samples \cite{Finn-ICML-2017}. The detailed implementation of MTAP can be found in Appendix C of \cite{Finn-ICML-2017}. Based on this setting, we obtained a probabilistic forecast model via an ensemble technique presented in \cite{Finn-ICML-2017} in the offline learning part. Specifically, we first sequentially trained $N_{off}$ individual probabilistic forecast models for $N_{off}$ offline forecast tasks with their corresponding historical data sets. Then, we parameterized the probabilistic forecast model with the average parameter vector across all individual models as $\theta_{MTAP}=\frac{1}{N_{off}}\sum_{i=1}^{N_{off}}\theta^{(i)}\text{, }$where $\theta^{(i)}$ is the parameter of the $ith$ individual probabilistic forecast model. The online mode of MTAP was also the same as our proposal. 

We hereinafter refer to our method as \emph{meta-learning} for short. Probabilistic forecast models for benchmarks and meta-learning all were built based on multiple LSTM layers with residual connections \cite{Meng-TPS-2022} to model temporal relations and quantiles were given by a fully connected layer. We denote the number of LSTM layers of the probabilistic forecast model as $N_L$ and set the size of hidden states for each layer as $H_L$.

    \begin{table}[!t]
      \renewcommand{\arraystretch}{1.3}
      \caption{Details of Tests $\mathscr{A}$}
      \vspace{-0.3cm}
      \label{table_testAB}
      \centering
      % \resizebox{\linewidth}{!}{
      \resizebox{6.5cm}{!}{
      \begin{tabular}{c|c}
      \hline\hline
      % & Test $\mathscr{A}$ \\ \hline
      Scenario & Lead time adjustment in online WPF \\ \hline
      \makecell*[c]{Location} 	&\makecell*[c]{\shortstack{Willogoleche Wind Farm, $P_c$ = 119.4MW}}  \\\hline
      \makecell*[c]{Covering period} 	&\makecell*[c]{\shortstack{$2018/12/31$ to $2020/12/31$}}  \\\hline
      Resolution 	&5 min  \\\hline
      \makecell*[c]{\shortstack{\shortstack{Input vector\\$\boldsymbol X_t$}}}	&\makecell*[c]{\shortstack{Historical wind power series of the wind\\ farm, time of the day, day of the year}} 
      \\\hline
      \makecell*[c]{Forecasting target\\$y_{t+\tau}$ in offline \\forecast tasks} 	&\makecell*[c]{\shortstack{Power of next 0.5, 1, 2, 4 h, respectively. Maximum, \\minimum, and average power during next 0.5, 1, 2,\\ 4 h, respectively. \emph{Offline forecast task 1 to 16}}} \\\hline
      \makecell*[c]{Forecasting target\\$y_{t+\tau}$ in online \\forecast tasks} 	&\makecell*[c]{\shortstack{Power of next 0.75, 1.5, 3 h, respectively. Maximum, \\minimum, and average power during next 0.75, 1.5, 3 h,\\ respectively. \emph{Online forecast task 1 to 12}}}  \\%\hline
      %Lead time  &5 minutes	&5 minutes	&One day  \\
      \hline\hline
      \end{tabular}}
      \vspace{-0.7cm}
      \end{table}

\subsubsection {Description of Data Sets and Tests}
We implemented a real-world data-based numerical test, i.e., test $\mathscr{A}$, which is detailed in Table~\ref{table_testAB}. The data set of test $\mathscr{A}$ is from the Australian National Electricity Market \cite{Wind-farm-data}. Detailed information for $y_{t+\tau}$ in different offline and online forecast tasks and features involved in $\boldsymbol X_t$ are also listed in Table~\ref{table_testAB}. Apart from wind power series, test $\mathscr{A}$ takes \emph{time of the day} and \emph{day of the year} as extra features in $\boldsymbol X_t$ to consider the diurnal and seasonal effects as mentioned in subsection IV-B of \cite{Hu-TNLS-2020}. 
% Denoting $N_s$ as the cumulative number of seconds countered from 00:00 to the time stamp of $y_{t+\tau}$, time of the day can be expressed as a data pair $\left\{\text{sin}\left(\frac{2\pi N_s}{24\times3600}\right)\text{, cos}\left(\frac{2\pi N_s}{24\times3600}\right)\right\}$. Similarly, day of the year is denoted as the data pair by replacing $N_s$ with the cumulative number of the day countered from January 1st to the day of $y_{t+\tau}$ and $24\times3600$ with the number of days in the year.% 
All features in $\boldsymbol X_t$ as well as $y_{t+\tau}$ have been normalized via min-max normalization before the forecast.

% \vspace{-0.2cm}
\subsubsection {Model Training and Hyperparameter Tuning}
The first 40\% samples of the data set were used for training (offline learning part), the middle 20\% for validation (offline learning part), and the last 40\% for testing (online learning part). A two-step grid-search method was adopted to determine the structure of the forecast model (associated with $N_L$ and $H_L$), the lag interval $\delta$, total number of inner loop updates $M$, sliding-window size $N_{\lambda}$, forgetting factor $\lambda$, iteration time of the incremental online learning $N_{inc}$, and the learning rates $\alpha_{in}$, $\alpha_{out}$, $\alpha_{on}$. Specifically, $N_L$ was chosen from \{8, 16, 32, 64\}, $W_L$ from \{16, 32, 64, 128\}, $\delta$ from \{2 h, 4 h, 8 h, 16 h\}, $M$ from \{1, 2, 4, 6\}, $N_{\lambda}$ from \{1, 2, 3, 4\}, $\lambda$ from \{0.2, 0.4, 0.6, 0.8\}, $N_{inc}$ from \{1, 2, 4, 6\}, and $\alpha_{in}$, $\alpha_{out}$, $\alpha_{on}$ from \{1e-3, 5e-3, 1e-2\}. First, the optimal combination of \{$N_L$, $H_L$, $\delta$, $M$, $\alpha_{in}$, $\alpha_{out}$\} was determined when the average meta-loss on the validation data set was the lowest during the grid-search process, which was recorded as \{16, 64, 8 h, 4, 5e-3, 1e-3\}. Later, we formulated a ``task stream'' based on the validation data set and offline forecast tasks with a predetermined $t_{T}$=0.5 h. The optimal set of \{$N_{\lambda}$, $\lambda$, $N_{inc}$, $\alpha_{on}$\} was determined by minimizing the average loss in the ``task stream'' for validation with incremental online learning via grid-search, which was recorded as \{3, 0.4, 4, 1e-3\}. The training of the model was implemented on CentOS 7.6 with 8 TITAN V GPUs and Adam was chosen for gradient descent.

\begin{table*}[!t]
  \renewcommand{\arraystretch}{1.3}
  \caption{Evaluation Metrics for Probabilistic WPF With Different Lead Times in Test $\mathscr{A}$}
  \vspace{-0.2cm}
  \label{table_performance_testA}
  \centering
  % \resizebox{\linewidth}{!}{
    \resizebox{13cm}{!}{
  \begin{tabular}{c|c|c|c|c|c|c|c|c|c|c|c|c}%
  \hline\hline
  % \cline{2-16}
  \multicolumn{1}{c|}{} &\multicolumn{3}{c|}{Average deviation $\overline{b_\tau}\%$} &\multicolumn{3}{c|}{Average PI width $\overline{\delta_\tau}$} &\multicolumn{3}{c|}{Average skill score $\overline{S_\tau}$} &\multicolumn{3}{c}{Mean absolute error MAE}
  \\\hline
    &\makecell*[c]{$t_{T}$=0.5 h} &\makecell*[c]{$t_{T}$=4 h} &\makecell*[c]{$t_{T}$=8 h} &\makecell*[c]{$t_{T}$=0.5 h} &\makecell*[c]{$t_{T}$=4 h} &\makecell*[c]{$t_{T}$=8 h} &\makecell*[c]{$t_{T}$=0.5 h} &\makecell*[c]{$t_{T}$=4 h} &\makecell*[c]{$t_{T}$=8 h} &\makecell*[c]{$t_{T}$=0.5 h} &\makecell*[c]{$t_{T}$=4 h} &\makecell*[c]{$t_{T}$=8 h} \\ \hline
  \makecell*[c]{Single-task\\learning}  
  &$4.70$ &$3.30$ &$3.68$ & $0.0558$ &$0.0260$ &$0.0182$ & $-0.222$ &$-0.149$ &$-0.106$ &$0.0284$ &$0.0182$ &$0.0134$ \\ \hline
  \makecell*[c]{MTAO}  
  &$1.98$ &$3.27$ &$3.32$ & $0.0488$ &$0.0230$ &$0.0180$ & $-0.194$ &$-0.119$ &$-0.091$ &$0.0235$ &$0.0156$ &$0.0121$  \\ \hline
  \makecell*[c]{MTAP}  
  &$3.14$  &$3.74$ &$3.47$ & $\boldsymbol{0.0384}$ &$\boldsymbol{0.0145}$ &$\boldsymbol{0.0147}$ & $-0.251$ &$-0.127$ &$-0.099$ &$0.0318$ &$0.0152$ &$0.0126$ \\ \hline
  \makecell*[c]{Meta-\\learning}  
  &$\boldsymbol{1.52}$   &$\boldsymbol{2.59}$ &$\boldsymbol{2.64}$ & $0.0444$ &$0.0180$ &$0.0153$ & $\boldsymbol{-0.160}$ &$\boldsymbol{-0.109}$ &$\boldsymbol{-0.079}$ & $\boldsymbol{0.0209}$ &$\boldsymbol{0.0139}$ &$\boldsymbol{0.0103}$ \\
  \hline\hline
    \end{tabular}}
    \vspace{-0.6cm}
  \end{table*}

\vspace{-0.5cm}
\subsection {Experimental Results}
\vspace{-0.1cm}
The evaluation metrics provided in section \uppercase\expandafter{\romannumeral2}-C were applied to the validation of temporal adaptations and only results for testing were recorded. Specifically, average deviation $\overline{b_\tau}$ in (\ref{b_tau}), average PI width $\overline{\delta_\tau}$ in (\ref{d_tau}), and the average skill score $\overline{S_\tau}$ in (\ref{s_tau}) were used to evaluate the probabilistic forecasting results. Mean absolute error MAE in (\ref{mae}) was used to evaluate the improved point forecasting results. Table~\ref{table_performance_testA} shows evaluation results under different task switching frequencies online (determined by $t_{T}$ in the ``task stream'') for test $\mathscr{A}$. From Table~\ref{table_performance_testA} we can see that meta-learning possesses the lowest deviation $\overline{b_\tau}$ (1.52\%) in the reliability evaluation when $t_{T}$ is 0.5 h (i.e., we change the forecast task every six time spots), which indicates the highest reliability. As for the sharpness evaluation, MTAO, MTAP, and meta-learning exhibit narrower average PI widths (measured by $\overline{\delta_\tau}$) than single-task learning. For the comprehensive evaluation of skill scores, single-task learning, MTAO, and meta-learning show higher scores than MTAP, and meta-learning is observed as the best one with the highest $\overline{S_\tau}$ ($-$0.160) among all methods. For the accuracy of the 0.5-$th$ quantile prediction, meta-learning still has obtained the lowest MAE showing the highest accuracy in the improved point forecasting. When we set $t_{T}$ as 4 h or 8 h, similar to the evaluation when $t_{T}$ is 0.5 h: Meta-learning is still the best associated with skill score and reliability with narrow PI widths, and it showcases prominent comprehensive performance in the improved point forecasting.

% Same as test $\mathscr{A}$, evaluation metrics for provincial wind generation predictions associated with different $t_{T}$ are presented in Table~\ref{table_performance_testB} for Test $\mathscr{B}$. With the highest switching frequencies of forecast tasks, i.e., $t_{T}$=1.5 h, MTAP provides the least $\overline{\delta_\tau}$ in the sharpness evaluation, but shows much higher deviations in the reliability evaluation (with $\overline{b_\tau}$ as 7.82\%) than other methods. Meta-learning exhibits the best overall performance on both the probabilistic and point evaluation metrics by holding the lowest deviations in the reliability evaluation (with $\overline{b_\tau}$ as 1.60\%), 0.5-$th$ quantile forecasting (with MAE and sMAPE as 0.0088 and 6.60\%, respectively), and the highest skill score ($-$0.064). When $t_{T}$ is 12 h or 24 h, although the difference between evaluation metrics of different methods is smaller than that when $t_{T}$ is 1.5 h, meta-learning is still superior over benchmarks considering the overall performance.

Summarizing the analyses for probabilistic forecasting with different task switching frequencies determined by $t_{T}$, %one can see that MTAP tends to achieve lower average widths of PIs in the sharpness evaluation (from columns 5 to 7 in Table~\ref{table_performance_testA}) while sacrificing reliability (from columns 2 to 4 in Table~\ref{table_performance_testA}). This results in inferior skill scores of MTAP compared with MTAO and meta-learning (sometimes, MTAP also offers a worse average skill score than that of single-task learning) as presented in columns 8 to 10 in Table~\ref{table_performance_testA}. On the whole, 
meta-learning demonstrates the most preferable performance on the probabilistic WPF associated with reliability (from columns 2 to 4 in Table~\ref{table_performance_testA}) and skill scores (from columns 8 to 10 in Table~\ref{table_performance_testA}) considering forecasting with different lead times under relatively narrow PI widths (from columns 5 to 7 in Table~\ref{table_performance_testA}) in all simulations. This benefits from the excellent few-shot-learning ability of meta-learning in temporal adaptations. For the improved point forecasting, we observe that higher skill scores generally incur lower MAE in the point forecasting results (from columns 11 to 13 in Table~\ref{table_performance_testA}). As meta-learning shows the highest skill score in the comprehensive evaluation of the probabilistic WPF, the high-quality point forecasting can be a by-product provided by our proposed two-part learning approach, which equips the flexibility of models to apply to both probabilistic and point forecasting. 

\begin{figure}
    \begin{center}
    \includegraphics[width=3.3in]{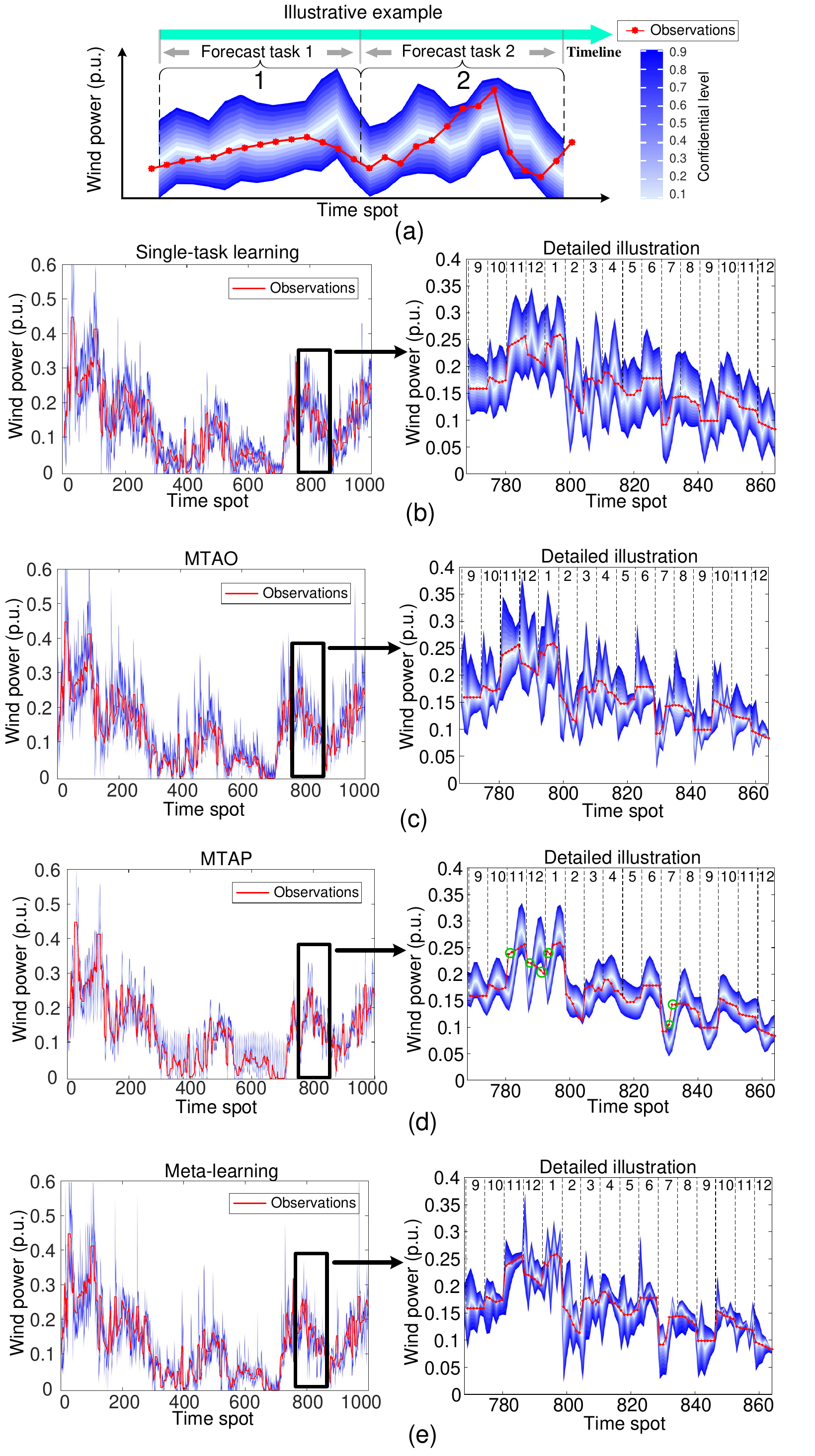}\\
    \vspace{-0.5cm}
    \caption{Probabilistic forecasting results of test $\mathscr{A}$ when $t_{T}$ is 0.5 h. (a) is an illustrative example of forecasting results of different forecast tasks implemented in the ``task stream''. (b), (c), (d), and (e) are forecasting results for single-task learning, MTAO, MTAP, and meta-learning, respectively.}\label{fan-chart}
    \end{center}
    \vspace{-0.9cm}
  \end{figure}

Probabilistic forecasting results (when $t_{T}$=0.5 h) reflecting confidential levels from 10\% to 90\% are presented in Fig.~\ref{fan-chart}. An illustrative example showing forecasting results in different forecast tasks online in the ``task stream'' is demonstrated in Fig.~\ref{fan-chart}(a), where the number separated by vertical dashed lines on the top denotes the label of different forecast tasks and probabilistic forecasting results between adjacent dashed lines were obtained by implementing the forecast task determined by the separated number. Fig.~\ref{fan-chart}(b), (c), (d), and (e) showcase probabilistic forecasting results of single-task learning, MTAO, MTAP, and meta-learning, respectively, where detailed results in all twelve online forecast tasks are demonstrated on the right side following the same way illustrated in the illustrative example. In each subfigure, the red line represents the observation of the forecasting target. The horizontal axis represents time spots for online forecasting in forecast tasks implemented in the ``task stream'' and the time interval between two successive forecasts in one forecast task is 5 min. Wider PI widths are witnessed in single-task learning and MTAO compared with those of meta-learning. The higher sharpness of meta-learning can be traced back to its prominent few-shot-learning ability necessitating higher adaptivity for different forecast tasks. Unreasonable narrow PI widths are demonstrated in MTAP, and some observations are out of the range of the 90\% confidential level in the detailed illustration of MTAP, which are marked with green circles in Fig.~\ref{fan-chart}(d). Among all tested results in MTAP and meta-learning, the proportion of observations outside the 90\% confidential level in MTAP is 26.3\%, which is much higher than that of meta-learning, i.e., 9.4\%. This leads to much lower reliability (column 2 in Table~\ref{table_performance_testA}) and skill score (column 8 in Table~\ref{table_performance_testA}) for MTAP compared with our proposal. Meanwhile, the average computation time for online forecasting (including generating online probabilistic WPF results and incremental online learning) at each time spot in meta-learning is 0.246 s, which is significantly shorter than lead times for online forecast tasks (from 0.75 h to 3 h).

In addition, we evaluated the average validation loss of forecast models after training with different amounts of samples for a randomly selected online forecast task. Two kinds of experiments were implemented by initializing the forecast model with meta-parameters and random initializations, respectively. Simulation results about training the forecast model with 5 epochs are shown in Table~\ref{sample_volume_time}. We can see that the forecast model initialized with meta-parameters, i.e., the base forecast model, obtains much lower loss than that of random initialization even when only a very limited amount of data for the new task is provided. Random initialization, even using a thousand times more data and ten times more training time, is not able to get a loss close to our proposal through several epochs. This validates the fast adaptation of our method. 

To sum up, meta-learning makes a great trade-off between sharpness and reliability and has achieved the best comprehensive performance on fast temporal adaptations.

\begin{table}[!t]
  \renewcommand{\arraystretch}{1.3}
  \caption{Comparison of Training Process}
  \vspace{-0.2cm}
  \label{sample_volume_time}
  \centering
  % \resizebox{\linewidth}{!}{
    \resizebox{9cm}{!}{
  \begin{tabular}{c|c|c|c|c|c|c|c|c}%
  \hline\hline
  % \cline{2-16}
  \multicolumn{1}{c|}{} &\multicolumn{4}{c|}{Average validation loss after 5 epochs} &\multicolumn{4}{c}{Average training time over 5 epochs}
  \\\hline
    &\makecell*[c]{$10$ \\samples} &\makecell*[c]{$100$ \\samples} &\makecell*[c]{$1000$ \\samples} &\makecell*[c]{$10000$ \\samples} &\makecell*[c]{$10$ \\samples} &\makecell*[c]{$100$ \\samples} &\makecell*[c]{$1000$ \\samples} &\makecell*[c]{$10000$ \\samples}\\ \hline
  \makecell*[c]{Meta-\\parameter}  
  &$0.0453$ &$0.0386$ &$0.0321$ & $0.0308$ &$6.01$ s &$6.68$ s &$12.0$ s &$69.8$ s\\ \hline
  \makecell*[c]{Random\\ initialization}  
  &$0.258$ &$0.255$ &$0.249$ & $0.231$ &$5.99$ s &$6.65$ s &$11.9$ s &$69.7$ s \\ 
  \hline\hline
    \end{tabular}}
    \vspace{-0.4cm}
  \end{table}

  \vspace{-0.3cm}

\section {Application \uppercase\expandafter{\romannumeral2}: Forecasting for the Newly Established Wind Farms}
\vspace{-0.1cm}
\subsection {Problem Description}
\vspace{-0.15cm}
Learning-based probabilistic forecast models generally exhibit satisfactory performance with sufficient historical data. However, a very limited amount of or even no historical data is available for newly established wind farms, thus severe overfitting may occur in the training process, deteriorating the efficacy of the forecast model (especially in deep learning). The probabilistic WPF for a newly established wind farm may be related to other wind farms in proximity to it. Typically, the proximity of wind farms can be measured by the distance{$\footnote{ Assuming the latitude and longitude for wind farm 1 are $\phi_1$ and $\lambda_1$, and those for wind farm 2 are $\phi_2$ and $\lambda_2$, the distance between these two wind farms can be estimated as $2\cdot atan2(\sqrt{a},\sqrt{1-a})\cdot R_e$, where $atan2(\cdot)$ is the 2-argument arctangent function, $a=sin\frac{(\phi_1-\phi_2)\pi}{360}+cos\frac{\phi_1\pi}{180}\cdot cos\frac{\phi_2\pi}{180}\cdot cos\frac{(\lambda_1-\lambda_2)\pi}{360}$, $R_e=6371km$ (the earth's mean radius).}$} between them, and wind farms within short distances can be defined as nearby wind farms. Nearby wind farms in relatively flat or uniform terrain generally share similar climate conditions, e.g., wind speed, relative humidity, temperature, etc. Therefore, we utilize information of nearby wind farms to help the probabilistic WPF of a new wind farm to address the above dilemma of data scarcity. Namely, we establish a base forecast model related to other nearby wind farms and apply it to the probabilistic WPF of the newly established ones. 

This application demonstrates the applicability of our method to newly established wind farms, i.e., the spatial adaptation. Besides, with the accumulation of historical data of the newly established wind farms, we can switch to Application \uppercase\expandafter{\romannumeral1} for the temporal adaptation of the probabilistic WPF.

\begin{table*}[!t]
  \renewcommand{\arraystretch}{1.3}
  \caption{Evaluation Metrics for Probabilistic WPF of the Newly Established Wind Farm in Test $\mathscr{B}$}
  \vspace{-0.3cm}
  \label{table_performance_testC}
  \centering
  % \resizebox{\linewidth}{!}{
    \resizebox{13cm}{!}{
      \begin{tabular}{c|c|c|c|c|c|c|c|c|c|c|c|c}%
        \hline\hline
        % \cline{2-16}
        \multicolumn{1}{c|}{} &\multicolumn{3}{c|}{Average deviation $\overline{b_\tau}\%$} &\multicolumn{3}{c|}{Average PI width $\overline{\delta_\tau}$} &\multicolumn{3}{c|}{Average skill score $\overline{S_\tau}$} &\multicolumn{3}{c}{Mean absolute error MAE} \\ \hline
        &\makecell*[c]{$\tau$=0.75 h} &\makecell*[c]{$\tau$=1.5 h} &\makecell*[c]{$\tau$=3 h} &\makecell*[c]{$\tau$=0.75 h} &\makecell*[c]{$\tau$=1.5 h} &\makecell*[c]{$\tau$=3 h}&\makecell*[c]{$\tau$=0.75 h} &\makecell*[c]{$\tau$=1.5 h} &\makecell*[c]{$\tau$=3 h}&\makecell*[c]{$\tau$=0.75 h} &\makecell*[c]{$\tau$=1.5 h} &\makecell*[c]{$\tau$=3 h} \\ \hline
  \makecell*[c]{Single-task\\learning}  
  &$8.17$ &$7.78$ &$7.81$ & $0.0180$ &$0.0175$  &$0.0176$ &$-0.109$ &$-0.105$ &$-0.101$ &$0.0149$ &$0.0143$  &$0.0136$   \\ \hline
  \makecell*[c]{MTAO}  
  &$5.58$ &$5.43$ &$5.56$ & $0.0183$ &$0.0178$ &$0.0171$ & $-0.101$ &$-0.104$ &$-0.092$ &$0.0138$  &$0.0141$  &$0.0126$  \\ \hline
  \makecell*[c]{MTAP}  
  &$9.81$ &$12.1$ &$12.0$ & $\boldsymbol{0.0144}$ &$\boldsymbol{0.0162}$ &$\boldsymbol{0.0116}$ & $-0.112$ &$-0.129$ &$-0.112$ &$0.0147$  &$0.0167$ &$0.0144$   \\ \hline
  \makecell*[c]{Meta-\\learning}  
  &$\boldsymbol{4.42}$ &$\boldsymbol{4.02}$ &$\boldsymbol{4.78}$ & $0.0171$ &$0.0174$ &$0.0180$  & $\boldsymbol{-0.095}$  &$\boldsymbol{-0.094}$ &$\boldsymbol{-0.085}$ &$\boldsymbol{0.0124}$  &$\boldsymbol{0.0129}$ &$\boldsymbol{0.0118}$  \\
  \hline\hline
    \end{tabular}}
    \vspace{-0.4cm}
  \end{table*}
  
  \begin{figure*}[!t]
    \begin{center}
    \includegraphics[width=5.5in]{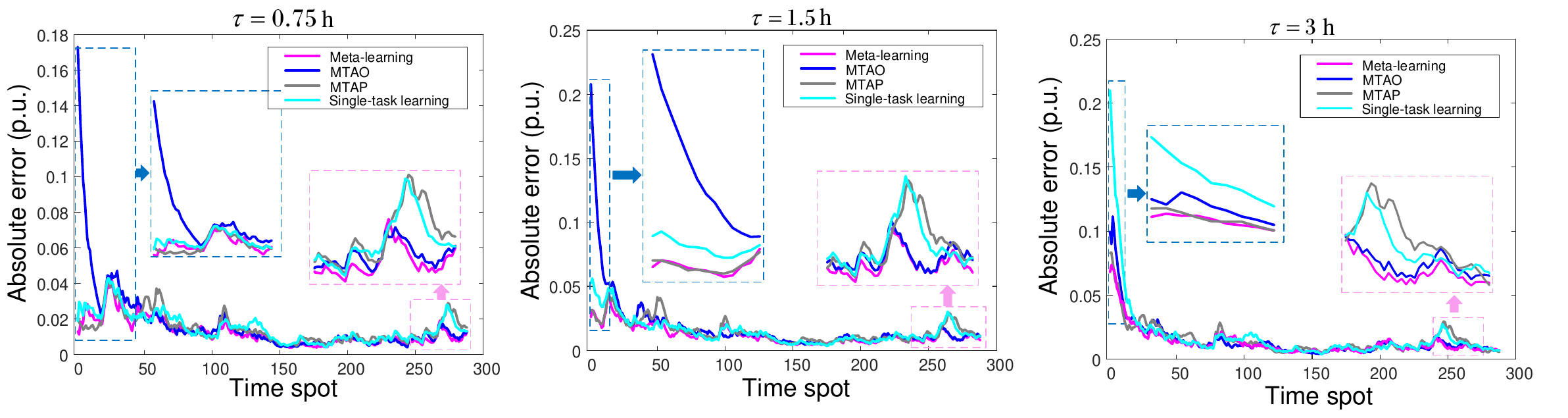}\\
    \vspace{-0.5cm}
    \caption{Absolute forecast errors in 0.5-$th$ quantile forecasting with different lead times for the first day after the wind farm was operated online.}\label{MAE}
    \vspace{-0.1cm}
    \end{center}
    \vspace{-0.7cm}
  \end{figure*}
  % \vspace{-1cm}

\vspace{-0.51cm}
\subsection {Application Method}
\vspace{-0.1cm}
Our proposed two-part learning approach in section \uppercase\expandafter{\romannumeral3}-B and \uppercase\expandafter{\romannumeral3}-C can also be applied to this application, where wind power outputs for different wind farms are referred to as different forecast tasks. In the offline learning part, the base forecast model is trained with mini-batch sets sampled from all generated data sets corresponding to nearby wind farms via inner and outer loops shown in Fig.~\ref{Offline part}. In the inner loop, we adapt the base forecast model to forecast tasks associated with different wind farms via (\ref{r-inner-loop}). In the outer loop, a meta-loss is formulated based on updating results in the inner loop via (\ref{r-meta-loss}) and minimized via (\ref{orig-grad}), where the loss calculated from different forecast tasks has the same weight. This is because we assume that information from each nearby wind farm contributes equally to the probabilistic WPF of the newly established ones. The training process is terminated with the same early stopping strategy in Application I. In this way, we endow the base forecast model with across-task knowledge of different nearby wind farms.

In the online learning part, the ``task stream'' only contains one online forecast task, i.e., probabilistic WPF of the newly established wind farms with a specified lead time. This task will not last long because we expect to evaluate the performance of forecasts under a small number of observations (at the very beginning of the operation of the newly established wind farms). Then, the trained base forecast model from the offline learning part is applied in the ``task stream'' combined with incremental online learning via (\ref{loss-on}) and (\ref{update-on}) for probabilistic WPF of the newly established wind farms. Since we only have one online forecast task here, the relationship between probabilistic forecast model at two successive online time spots is the same as that under \emph{condition 1} presented in subsection A of Application I.

\begin{table}[!t]
  \renewcommand{\arraystretch}{1.3}
  \caption{Details of Tests $\mathscr{B}$}
  \vspace{-0.3cm}
  \label{table_testC}
  \centering
  % \resizebox{\linewidth}{!}{
  \resizebox{6.5cm}{!}{
  \begin{tabular}{c|c}
  \hline\hline
  % & Test $\mathscr{B}$ \\ \hline
  Scenario & WPF for newly established wind farm \\ \hline
  \makecell*[c]{Location} 	&\makecell*[c]{\shortstack{Brown Hill Wind Farm, $P_c$ = 94.5MW \\Hallett Hill Wind Farm, $P_c$ = 71.4MW\\North Brown Hill Wind Farm, $P_c$ = 132.3MW\\Bluff Range Wind Farm, $P_c$ = 52.5MW\\Willogoleche Wind Farm, $P_c$ = 119.4MW}}  \\\hline
  \makecell*[c]{Covering period} 	&\makecell*[c]{\shortstack{$2018/12/31$ to $2020/12/31$}}  \\\hline
  Resolution 	&5 min  \\\hline
  \makecell*[c]{\shortstack{Input vector\\{$\boldsymbol X_t$}}}	&\makecell*[c]{\shortstack{Historical wind power series of the five wind\\ farms, time of the day, day of the year}} 
  \\\hline
  \makecell*[c]{$y_{t+\tau}$ in offline \\forecast tasks} 	&\makecell*[c]{\shortstack{Power of next 0.5, 1, 2, 4 h for four of the five wind \\farms, respectively. \emph{Offline forecast task 1 to 16}}} \\\hline
  \makecell*[c]{$y_{t+\tau}$ in online \\forecast tasks} 	&\makecell*[c]{\shortstack{Power of next 0.75, 1.5, 3 h for the rest one wind\\ farm, respectively. \emph{Online forecast task 1 to 3}}}  \\%\hline
  %Lead time  &5 minutes	&5 minutes	&One day  \\
  \hline\hline
  \end{tabular}}
  \vspace{-0.7cm}
  \end{table}

\vspace{-0.5cm}
\subsection {Setting of Numerical Simulations}
\vspace{-0.1cm}
As in Application I, we also set single-task learning, MTAO, and MTAP as benchmarks. A test $\mathscr{B}$ was implemented based on data sets of five nearby wind farms from the Australian National Electricity Market \cite{Wind-farm-data}, of which the detailed information is presented in Table~\ref{table_testC}. These wind farms are located at similar altitudes (around 600m height above sea level) in a relatively flat area in South Australia, and the maximum and minimum distances between different wind farms are 33.0km and 1.9km, respectively. We randomly chose one of these five wind farms as the newly established one. Historical samples with time stamps from 2018/12/31 to 2020/12/30 of the other four wind farms were used for training (80\%) and validation (20\%) in the offline learning part, and samples with time stamps from 2020/12/30 to 2020/12/31 (in one day) for the newly established one were used for testing in the online learning part. The same two-step grid-search method was adopted as in Application I to find the optimal combination of \{$N_L$, $H_L$, $\delta$, $M$, $N_{\lambda}$, $\lambda$, $N_{inc}$, $\alpha_{in}$, $\alpha_{out}$, $\alpha_{on}$\}, which was determined as \{16, 64, 8 h, 4, 3, 0.6, 4, 5e-3, 1e-3, 1e-3\}.

\vspace{-0.3cm}
\subsection {Experimental Results}
\vspace{-0.1cm}

For the validation of the applicability of our method to the newly established wind farms, the average deviation $\overline{b_\tau}$ (for reliability), average PI width $\overline{\delta_\tau}$ (for sharpness), average skill score $\overline{S_\tau}$ (for comprehensive evaluation of both reliability and sharpness), and mean absolute error MAE (for improved point forecast) evaluated in Application I were also used for the evaluation of probabilistic and point forecasting results here. Detailed testing results of these metrics about forecasts with different lead times $\tau$ for the newly established wind farm are displayed in Table~\ref{table_performance_testC} for test $\mathscr{B}$. Similar to results in test $\mathscr{A}$, 
%it is also noticed that MTAP offers the narrowest PI widths in the sharpness evaluation (the smallest $\overline{\delta_\tau}$) in all experiments. This leads to inferior performance on $\overline{b_\tau}$, $\overline{S_\tau}$, MAE, and sMAPE for MTAP compared with the other multi-task learning method, i.e., MTAO, owing to unreasonably narrow PI widths estimated by MTAP. %MTAO still exhibits higher overall performance than single-task learning benefiting from richer information of data about multiple forecast tasks.%
meta-learning demonstrates the highest reliability (lowest $\overline{b_\tau}$ shown from columns 2 to 4 in Table~\ref{table_performance_testC}), skill score (highest $\overline{S_\tau}$ shown from columns 8 to 10 in Table~\ref{table_performance_testC}), and accuracy for improved point forecasting (lowest MAE shown from columns 11 to 13 in Table~\ref{table_performance_testC}) in test $\mathscr{B}$ owing to its great few-shot-learning ability. As a whole, considering the comprehensive evaluation with an appropriate estimate of the PI widths, simulation results in Table~\ref{table_performance_testC} showcase the greatest superiority of meta-learning in both probabilistic and point forecasting. These analyses indicate the highest applicability of our method to the newly established wind farms.
%Moreover, these experiments with different lead times also elucidate the efficacy of meta-learning for the temporal adaptation if forecast tasks of nearby wind farms consider different lead times.

Based on the above analyses, it can be inferred that the main conclusion drawn from probabilistic forecasting results with different confidential levels in this application will be similar to those visualized in Fig.~\ref{fan-chart} in Application I. Thus, we do not present probabilistic WPF results here considering the space limitation. Instead, detailed evaluation results$\footnote{\vspace{-1cm}These results are smoothed for clearer visualizations.}$ of MAE for different experiments with different lead times in test $\mathscr{B}$ are placed in Fig.~\ref{MAE}. In our setting, this simulation demonstrates the absolute forecast error for the first day after the wind farm was operated online, and the horizontal axis in each figure represents successive time spots with 5-min resolution. We see that meta-learning introduces no significantly large deviation all the time. It offers lower absolute errors at the beginning of the online forecasting and also presents lower errors at the end of the day, thus lower MAE is achieved by meta-learning in Table~\ref{table_performance_testC} compared with benchmarks. For the average computation time related to online forecasting with incremental online learning at each time spot in meta-learning, it is 0.257 s and is much shorter than the lead times for online forecast tasks, i.e., 0.75 h, 1.5 h, and 3 h.

In summary, meta-learning effectively utilizes information about nearby wind farms and demonstrates the greatest spatial adaptations. 
%at the beginning of the operation of the newly established wind farms.

\vspace{-0.4cm}
\section{Conclusion}

A two-part learning approach for probabilistic WPF has been designed and applied to applications concerning temporal and spatial adaptabilities in this paper. In our presented method, the combination of online learning with meta-learning exhibits significant positive effects benefiting from the excellent few-shot-learning ability of meta-learning. The main advantage of our approach is the prominent adaptability of probabilistic forecasts for different online forecast tasks based on small amounts of online data. Numerical simulations corroborate the superiority of the proposed method over benchmarks considering the accordance with reality, PI width, and 0.5-$th$ quantile forecasting.

% or


% ============================================
%\appendices
%Appendix one text goes here %\cite{Roberg2010}.

%\section{}

%\section*{Acknowledgment}

%The authors would like to thank D. Root for the loan of the SWAP. The SWAP that can ONLY be usefull in Boulder...

\ifCLASSOPTIONcaptionsoff
  \newpage
\fi

\vfill


\end{document}